\documentclass[twocolumn]{aastex63}
\received{October 30, 2019}
\revised{December 13, 2019}
\accepted{January 7, 2020}
\shorttitle{Detrending transit LCs with LSTMs}
\shortauthors{Morvan et al.}
\graphicspath{{./}{figures/}}

\begin{document}

\title{Detrending Exoplanetary Transit Light Curves with Long Short-Term Memory Networks}

\email{mario.morvan.18@ucl.ac.uk, n.nikolaou@ucl.ac.uk\\ angelos.tsiaras.14@ucl.ac.uk, ingo.star@ucl.ac.uk}

\author{Mario Morvan}
\affiliation{Department of Physics and Astronomy, University College London, Gower Street, London, WC1E 6BT, UK}

\author{Nikolaos Nikolaou}
\affiliation{Department of Physics and Astronomy, University College London, Gower Street, London, WC1E 6BT, UK}

\author{Angelos Tsiaras}
\affiliation{Department of Physics and Astronomy, University College London, Gower Street, London, WC1E 6BT, UK}

\author{Ingo P. Waldmann}
\affiliation{Department of  Physics and Astronomy, University College London, Gower Street, London, WC1E 6BT, UK}



\begin{abstract}

The precise derivation of transit depths from transit light curves is a key component for measuring exoplanet transit spectra, and henceforth for the study of exoplanet atmospheres. However, it is still deeply affected by various kinds of systematic errors and noise. In this paper we propose a new detrending method by reconstructing the stellar flux baseline during transit time. We train a probabilistic Long Short-Term Memory (LSTM) network to predict the next data point of the light curve during the out-of-transit, and use this model to reconstruct a transit-free light curve -- i.e. including only the systematics -- during the in-transit. By making no assumption about the instrument, and using only the transit ephemeris, this provides a general way to correct the systematics and perform a subsequent transit fit. The name of the proposed model is TLCD-LSTM, standing for \emph{Transit Light Curve Detrending LSTM}. Here we present the first results on data from six transit observations of HD\,189733\,b with the IRAC camera on board the \textit{Spitzer Space Telescope}, and discuss some of its possible further applications.
\end{abstract}



\section{Introduction} \label{sec:intro}

Since the first exoplanet atmosphere observation twenty years ago \citep{charbonneau_detection_2000}, more than 3000 transiting extrasolar planets have been discovered. 
Transit spectroscopy - i.e. multi-wavelength transit observations - has opened the way for the characterization of atmospheric content and properties of exoplanets. In effect, this can be done by first reconstructing the transmission or emission spectrum from the transit depth measurements at various wavelengths, and at a typical precision level of just a few parts-per-million (ppm) for hot gaseous planets. This is to be contrasted with the imprints left in the stellar light curve by various instrumental and astrophysical effects which make the measurement of the transit depths extremely challenging. Given the shift of the field towards increasingly smaller planets, the need for efficient detrending methods is thus ever growing. Here we present a long- short term memory (LSTM) neural network approach to effectively model and detrend instrument and astrophysical systematics in transit light curves.\\
The total flux $F(t)$ received by a detector at time $t$ can be broken down as follows:
\begin{enumerate}
    \item Star flux: $F_s(t)$
    \item Planetary signal: $\delta(t) = (R_P(t)/R_S)^2$ in the case of primary transit obstruction with no limb darkening, where $R_S$ is the stellar radius and $R_P$ the apparent planetary radius 
    \item Background stars and transient events: $F_b(t)$
    \item Noise and instrumental systematics: $G(.)$
    
\end{enumerate}
The total flux received by each pixel of the detector can then be written as $F(t) = G\big( (1-\delta(t))F_s(t) + F_b(t)\big)$, where $F_s$ and $F_b$ may vary depending on the position on the detector and are then subject to instrumental systematics. We will refer to individual pixel time series as \emph{pixel light curves}, and to the summed contribution of pixels over time as a \emph{raw light curve}. 

Essentially, the main instrumental systematics trend observed both with the Hubble WFC3 and the Spitzer IRAC cameras are the so-called \textit{ramp} effect \citep{knutson_map_2007}, hypothesized to be due to the charge trapping in the detector \citep{agol_climate_2010}, and intra-pixel and inter-pixel variations which are correlated with the position of the source on the detector which shows variations in quantum efficiency across different pixels\footnote{This effect has been described in the IRAC instrument handbook: \url{http://irsa.ipac.caltech.edu/data/SPITZER/docs/irac}}. 

Footprints of these entangled variability sources can be found in additional instrumental data collected besides the detector raw flux. In particular, the center and scale of the stellar point spread function (PSF) can be processed to give valuable information on the systematics while being mostly uncorrelated with the planetary signal itself.

Considering the analysis of time-correlated light curves with the end goal of detrending transit light curves and extracting the transit parameters as precisely as possible, one can approach the problem in several ways. 
Indeed, the disentanglement of various independent signals might naturally guide one toward \emph{blind source separation} techniques, which have been applied on this problem (\citealp{waldmann_cocktail_2012}, \citealp{morello_new_2014}, \citealp{morello_repeatability_2016}) using the pixel light curves as correlated components. In a complementary way, signal processing analysis techniques have also been used to denoise the raw or pixel light curves, with \emph{Gaussian processes} \citep{gibson_gaussian_2012}, \emph{pixel level decorrelation} \citep{deming_spitzer_2015} or \emph{wavelet analysis} (\citealp{carter_parameter_2009}, \citealp{thatte_selective_2010}, \citealp{morello_repeatability_2016}). 
Here we choose the angle of \emph{interpolation}, i.e. we want to provide predictions for the raw light curve during the transit time provided the out-of transit parts of the light curves. The interpolation method we propose is non-linear and thus capable of capturing complex long term dependencies in the light curve.

The use of \emph{artificial neural networks} (ANNs) is burgeoning in various fields including Astronomy. In particular, \cite{charnock_deep_2017} presented one of the first use of \emph{recurrent neural networks} (RNNs) in astronomy for supernovae classification. Yet, in the subfield of exoplanetary sciences, only a few studies have been using ANNs so far, with namely \cite{hinners_machine_2018} who predicted stellar and planetary parameters from Kepler light curves using RNNs and representation learning, \cite{zingales_exogan:_2018} on exoplanetary spectra retrieval, \cite{shallue_identifying_2018}, \cite{ansdell_scientific_2018}, \cite{osborn_rapid_2019} for the supervised classification of transit candidates and \cite{gomez_gonzalez_supervised_2018}, \cite{yip_pushing_2019} for planet detection in direct imaging. 

Here we make use of a \emph{long short-term memory (LSTM) neural network} \citep{hochreiter_long_1997} to interpolate the flux of a raw light curve during the transit, given additional time-series data coming from the PSF centroid. The LSTM network learns to predict the next value of the light curve at each time step. The predictions of future time steps are then performed in a probabilistic manner using \emph{ancestral sampling}, i.e. by injecting the current prediction as input to the subsequent prediction and so on.  We thus assume that the pre-transit and post-transit information, along with additional data such as centroid time-series, are sufficient to predict the flux that the detector would have received in the absence of a planet transit.

This paper is organised as follows: Section~\ref{sec:nn} contains 
background information about neural networks, Section~\ref{sec:method} presents the interpolating model and how it can be used for transit light curve fitting, and finally Section~\ref{sec:appli} is dedicated to an application on Spitzer data. 

\section{Recurrent Networks and LSTMs} \label{sec:nn}


In a typical \emph{supervised statistical learning} task, the goal is to learn a model $h(x) \simeq y$ that maps an input $x$ to an output $y$\footnote{Note that $x$ and $y$ can be scalars, or more generally $n$-dimensional vectors.} given examples of pairs $(x,y)$ in such a way that the expected error of future predictions is minimized. 

\begin{figure}
\begin{center}
\includegraphics[width=\columnwidth]{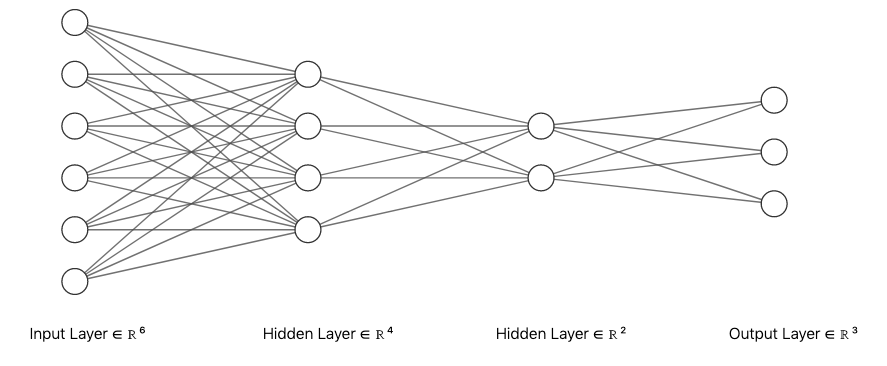}
\caption{An example of a feed-forward neural network with 2 hidden layers. When evaluating the underlying function $h$, the information is flowing from the input on the left towards the output layer on the right.  \label{fig:nn}}
\end{center}
\end{figure}

\textit{Feed-forward neural networks} or \emph{multi-layer perceptrons (MLPs)} represent the simplest architecture of deep neural networks. An example of this type of architecture is shown on Figure \ref{fig:nn}.
No feedback connections exist in these models.
Every layer consists of a set of neurons and the neurons of the input layer represent each of the original input variables $x$. The output of each neuron is a scalar value and is used as input for the neurons of the next layer. Each subsequent layer transforms a linear co mbination of the outputs of the neurons of the previous layer using an \emph{activation function} $\sigma$: $h_{l+1} = \sigma(W_lh_{l} +b_l)$ where $W_l$ is a matrix of multiplicative weights, $b_l$ the bias vector, $h_{l}$ the vector of units and $\sigma_l$ the \emph{activation function}, all at layer $l$. If we interchangeably  write  $h_l$ for the function represented at layer $l$ as well as its output, the full function represented by a feed-forward network can then be written: $y = h_D(h_{D-1}(...h_1(X)))$ where $D$ is the \emph{depth} of the network. Note that the non-linearity of at least one of the layers activation functions is key to obtaining a non-linear predictor.

\begin{figure}
\begin{center}
\includegraphics[width=\columnwidth]{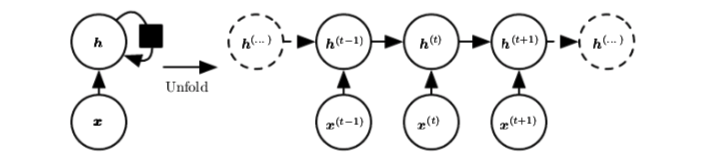}
\caption{An example of a recurrent neural network with no output from \citet{ian_goodfellow_deep_2016}\label{fig:rnn}.}
\end{center}
\end{figure}

The main characteristic of \emph{Recurrent} Neural Networks is that they allow for \emph{recurrent connections}\footnote{This means that -- unlike in feed-forward neural networks -- in RNNs the output of neurons from one layer can be used as input for neurons of the same or a previous layer.}. If we consider an input sequence $\{x_1, x_2...\}$ of vectors, a recurrent hidden layer will thus process it sequentially, receiving at step $t$ both the input $x_t$ as well as other previous \textit{hidden state(s)} in order to compute the current state $h_t$. A typical example is shown on Figure \ref{fig:rnn}, where the recurrence occurs between the hidden units of the same layer: $h_t=h_t(x_t,h_{t-1})$.
Compared to MLPs, RNNs allow us to reduce the number of parameters of the network by sharing weights between time-steps while seeking temporal patterns in the data. 

\begin{figure}
\begin{center}
\includegraphics[width=\columnwidth]{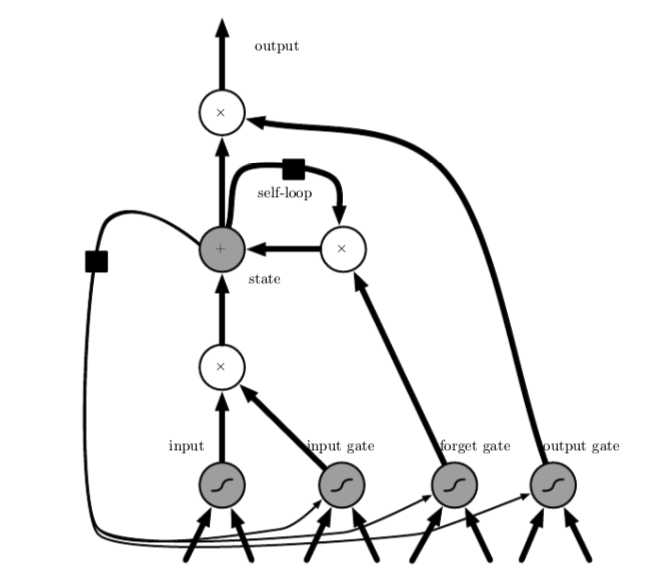}
\caption{An LSTM cell from \citet{ian_goodfellow_deep_2016}\label{fig:lstm}, which replaces a usual hidden unit (i.e. neuron) in a feed-forward neural network. The input, forget and output gating units enable the cell to accumulate or shut off respectively the current input, long-term dependencies and output through a sigmoidal activation function. The square here indicates a delay of one-time-step, and operation symbols in the circles indicate the operation involving the gates' outputs.}
\end{center}
\end{figure}

In practice, several more sophisticated recurrent architectures are often more effective than basic RNNs, with most being variants of the long short-term memory \citep[LSTM,][]{hochreiter_long_1997} architecture whose cell is shown in Figure \ref{fig:lstm}. LSTM networks have proven successful in a large range of applications including unconstrained handwriting recognition \citep{graves_novel_2009}, speech recognition \citep{graves_speech_2013}, machine translation \citep{sutskever_sequence_2014}, to cite only a few. An LSTM cell contains four different \emph{gates} (see Figure \ref{fig:lstm}), allowing the network to either retain or forget information from the past of the input sequence. This enables the relevant long-term time dependencies to be picked up more easily.  The main addition in LSTMs compared to the basic RNNs has been to introduce self-loops, which are conditioned on the context and controlled by the gates. Below we state the detailed update formulae for the gates and states composing each LSTM unit:
\begin{itemize}
    \item The input gate: $i_t = W_{ix}x_t + W_{ih}h_{t-1}+b_i $
    \item The forget gate: $f_t = W_{fx}x_t + W_{fh}h_{t-1}+b_f $
    \item The output gate: $ o_t = W_{ox}x_t + W_{oh}h_{t-1}+b_o $
    \item The cell state: $ c_t = \sigma(f_t) \odot c_{t-1} + \sigma(i_t) \odot \tanh(j_t)$
    \item The output vector: $ h_t = \sigma(o_t) \odot \tanh(c_t) $
\end{itemize}

Where $t$ denotes the time step, $W_{ab}$ the matrix of weights relative to the vectors $a$ and $b$, $b_a$ the bias vector relative to $a$, $\odot$ the Hadamart (i.e. entrywise) product and $\sigma$ is the activation function, typically a logistic sigmoid or $\tanh$ function.

Incidentally, these types of gated RNNs also have the advantage of being easier to train than basic RNNs, by alleviating the well known vanishing or exploding gradient issue~\footnote{Neural networks are trained via gradient-based minimization of a loss function. In each iteration of training, each parameter of the model (weight) receives an update proportional to the partial derivative of the loss w.r.t the current weight. Allowing these gradients to grow vanishingly small or too large can cause numerical instabilities, slow down training or stop it prematurely.}~\citep{kolen_gradient_2001}. 

\section{TLCD-LSTM} \label{sec:method}

Here we describe the proposed model to interpolate a time-series on a pre-defined prediction range. As the final goal of this paper is to study the transit signal contained in the interpolation range after correction of the systematic errors, we name the method \emph{Transit Light Curve Detrending LSTM (TLCD-LSTM)}.

The model is based on the deep auto-regressive neural network model described in \citet{salinas_deepar:_2017}. It assumes that temporal relations exist in the time-series and learns to predict the next step in the training range of the input time-series. It can also make use of additional data available for prediction contained in the so-called \emph{covariate time-series}, which is to be distinguished from the main time-series. In general, one can consider both the main and covariate time-series to be multi-variate, i.e. to be composed of several time-series each. 

TLCD-LSTM is specifically adapted for interpolation within a given range, and therefore differs from \citet{salinas_deepar:_2017} mainly in that the values it tries to predict are not in the future (i.e. the end of the time-series) but in timesteps somewhere within the time-series. 


\subsection{Model description} 
Let us denote with $\{x_1, x_2, ..,x_T\}$ (abbreviated $\{x_t\}$) the \emph{main time-series} of length $T$ we ought to interpolate on the prediction range $[t_1 .. t_2]$ with $t_1$ and $t_2$ integers in $[1..T]$, and $\{z_1, z_2, ..,z_T\}$ (abbreviated $\{z_t\}$) the time-series of covariates, which constitute additional data available for prediction on the whole time range. Finally, let us also denote with $\{y_1, y_2, ..,y_T\}$ (abbreviated $\{y_t\}$) the \emph{target time-series}, identical to the main time-series in the training range but which may differ in the prediction range. In the case of $\{x_t\}$ being a transit light curve, $\{y_t\}$ is the hypothetical light curve without any transit signal.

\begin{figure}[!ht]
\begin{center}
\includegraphics[width=\columnwidth]{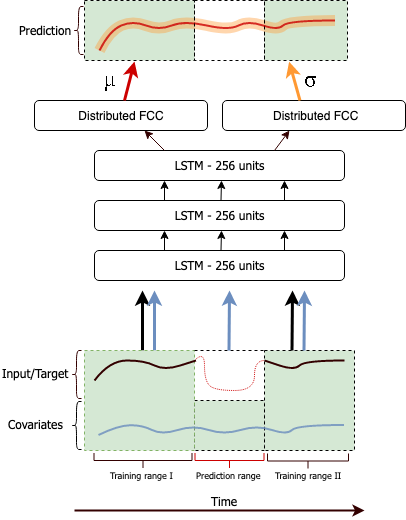}
\caption{Sketch of the interpolating probabilistic LSTM neural network. The main and covariate time-series are processed through three LSTM layers consisting of 256 units each, and then decoded into two outputs for each of the interpolated points: the mean and the standard deviation.\label{fig:archi}}
\end{center}
\end{figure}

As sketched in Figure \ref{fig:archi}, each value of the input time-series passes through a stack of LSTM layers, the output of which branches into two distinct feed-forward layers outputting two parameters $\widehat{\mu_t}$ and $\widehat{\sigma_t}$ at each time-step, which are the predicted mean and standard deviation for the distribution of the current value $x_t$, respectively. The details and hyperparameters of the architecture are presented in Appendix \ref{sec:hyperparameters}.
The same network is used both for the training and prediction ranges with only the inputs differing in each case.
\begin{table}[!ht]
    \centering
    \begin{tabular}{c|c|c|c}
        Mode & Range &  Inputs at $t$ & Output at $t$ \\
        \hline
        Training & $[1..t_1] \cup[t_2..T]$ & $x_{t-1}, z_t$  & $(\widehat{\mu_t}, \widehat{\sigma_t})$ \\
        Prediction & $[t_1..t_2]$ & $\widehat{\mu_{t-1}}, z_t$ & $(\widehat{\mu_t}, \widehat{\sigma_t})$ 
    \end{tabular}
    \caption{Differences of inputs at each time step $t$ between the training and prediction ranges. \label{tab:in-out}}
\end{table}

At each timestep $t$, the network predicts the current value $x_{t}$ from all past timesteps $x_1,..,x_{t-1}$ as well as from the current covariate $z_{t}$. While the actual previous time-series value $x_{t-1}$ is used as input in the training ranges, in the prediction range the previous prediction $\mu_{t-1}$ is injected as an input instead of it (see Table \ref{tab:in-out}).

\subsubsection{Training the model}
We assume each value $z_t$ is sampled from a normal distribution: $$z_t \sim \mathcal{N}(\mu,\,\sigma^{2})$$ 
The loss function is then computed as the product of individual likelihoods outside the prediction range: $$\mathcal{L}(\mu_{t_1..t_2}) = \prod_{t\in [1..t_1]\cup [t_2..T]} \frac{e^{(x_t-\widehat\mu_t)^2/2\widehat\sigma_t}}{\sqrt{2\pi\sigma_t}}$$
Note that the log-loss is only computed in the training ranges. However, the last output of the prediction range is taken as the first input of the second training range, thus providing a way to link together the outputs in the different ranges.

\subsubsection{Predicting the time-series}
There are several ways one can generate predictions in $[t_1..t_2]$, once a model is trained. Since the inputs of the network consist of parameters of a probability distribution, the simplest one is to directly take the vector of predicted means $\widehat{y_t} = \mu_t$. However, one can also generate a \textit{trace} by drawing every value from the Gaussian distribution at every timestep in the prediction range: $\widehat{y_t}\sim \mathcal{N}(\mu_t,\,\sigma_t^{2})$, and injecting each of these predictions as input for the next time step. Multiple traces obtained with this process then represent the joint predicted distribution (of which they are samples) in a more general way than merely using the means vector. To generate a single vector of predictions from multiple traces, one can -- for instance -- select the median or mean value at every timestep to construct the \textit{median trace} or \textit{mean trace} on the prediction range. In Section \ref{sec:appli}, we focus on the simplest approach, i.e. selecting the output means and standard deviations.

\subsubsection{Covariant Features}
The covariates time-series $\{z_t\}$ can consist of single-dimensional or multi-dimensional data available both in the training and prediction ranges. It is used by the network as additional information besides the target time-series. This works merely by concatenating $x_t$ (conversely $\widehat{x_t}$ in prediction mode) to the covariate data $z_t$ to construct the new input to the network at every timestep. Ideally, one wants $\{z_t\}$ to be correlated with the target time-series.
Several time-series might be related to the time-correlated noise we intend to correct, and therefore can be used as covariate data in the model. In the application presented in Section \ref{sec:appli} we suggest the use of PSF-related time-series, namely the instrument's point spread function (PSF) centers and widths of a 2D Gaussian fit on the images at every time step. One could also think of other potentially relevant information such as simultaneous host star activity, calibration data relative to the detector and estimations of background flux. For ground-based applications, information about airmass, seeing and weather patterns could be included.

\subsection{Application to transit light curves}
Here we discuss the use of the interpolating model specifically to transit light curves. 

The transit signal must be contained within the prediction range. This requires either to know beforehand when the transit occurs, or to adapt the prediction range during the first phase of the training. Pre-transit and post-transit data are used for training the network, and are assumed to not be contaminated by any transit event. They can however contain any sort of variability coming from the star, the background or the instrument. In fact, the model aims at picking up variations due to all sources other than a transit event in order to predict the flux due to these sources alone during the transit time.



We perform a transit fit at each evaluation step even though our model does not strictly require it for the training. This is done for two main reasons: 

1) The transit fit can be used as a proxy to evaluate the quality of the prediction and provide us with a criterion for early-stopping the training of our model. The transit fit is performed on the detrended light curve normalized with respect to the star $(1-\delta_t)$. For details, see Appendix \ref{ap:transit_fit}. 
 2) We can use the transit fit to adapt the prediction range $[t_1..t_2]$ during training so that it matches better the actual transit range of the data. This can be done by extracting the fitted mid-transit time and transit duration to compute the times for the beginning and end of transit.

\section{Application} \label{sec:appli}

We present an application to 6 transit observations of planet HD\,189733\,b from the Spitzer/IRAC detector at 8\,$\mu$m, collected in 2007 and 2008 (PI: E. Agol, program: 40238). This hot-Jupiter planet has been extensively studied and makes a good candidate for bench-marking our method. In this wavelength channel, the ramp effect can be heavily pronounced \citep{agol_climate_2010}, while the  intra-pixel variations due to pointing jitter are less important than at shorter wavelengths. A few preprocessing steps are applied to the data\footnote{The data used are publicly available and were downloaded from: \url{https://sha.ipac.caltech.edu/applications/Spitzer/SHA/}} and detailed in Appendix \ref{ap:preprocessing}. These include outlier removal, raw light curve extraction and normalization, centroids fitting and background light estimation. 

In this section, we present predictions on the pre-transit range (on intervals not used in the training of the model) as an initial evaluation of the model, and then show results on the real transit ranges on which we derive the detrended light curve and subsequent transit fit.

\begin{figure*}[t]
\begin{center}
\includegraphics[width=\textwidth]{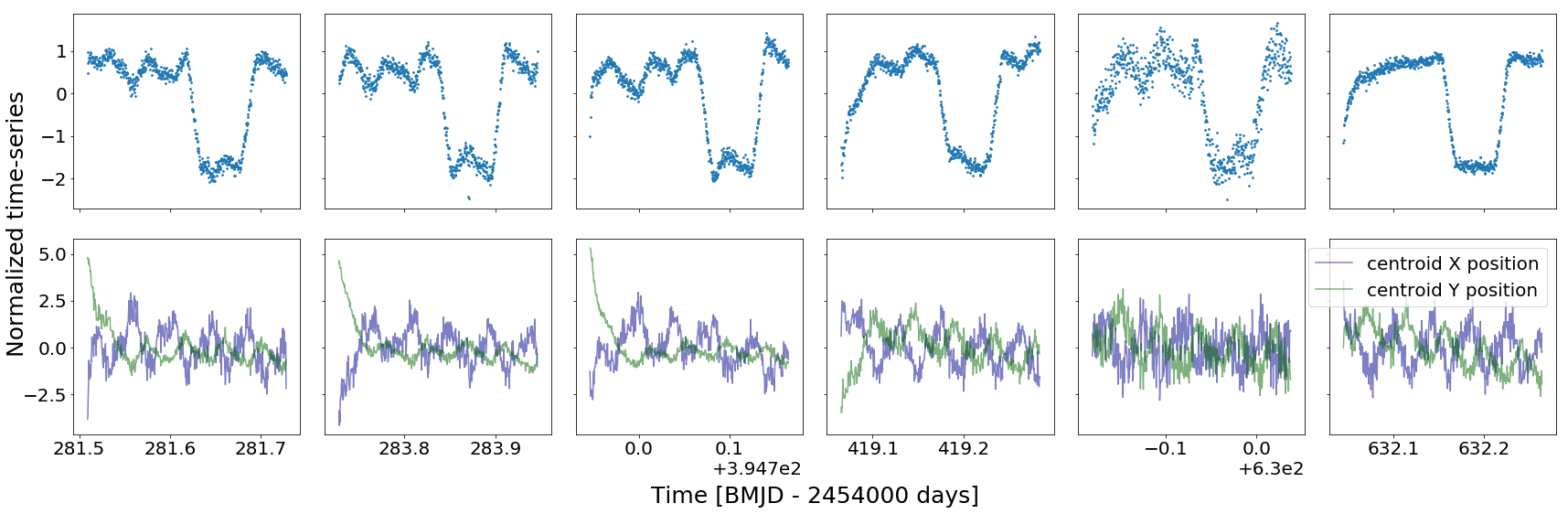}
\caption{ \textit{(Top)} 6 Spitzer/IRAC 8$\mu m$ raw transit light curves of HD\,189733\,b after preprocessing.  \textit{(Bottom)} X/Y centroid positions of the point spread function. \label{raw_data}}
\end{center}
\end{figure*}


\subsection{Testing \label{subsec:testing}}
\begin{figure}
\begin{center}
\includegraphics[width=\columnwidth]{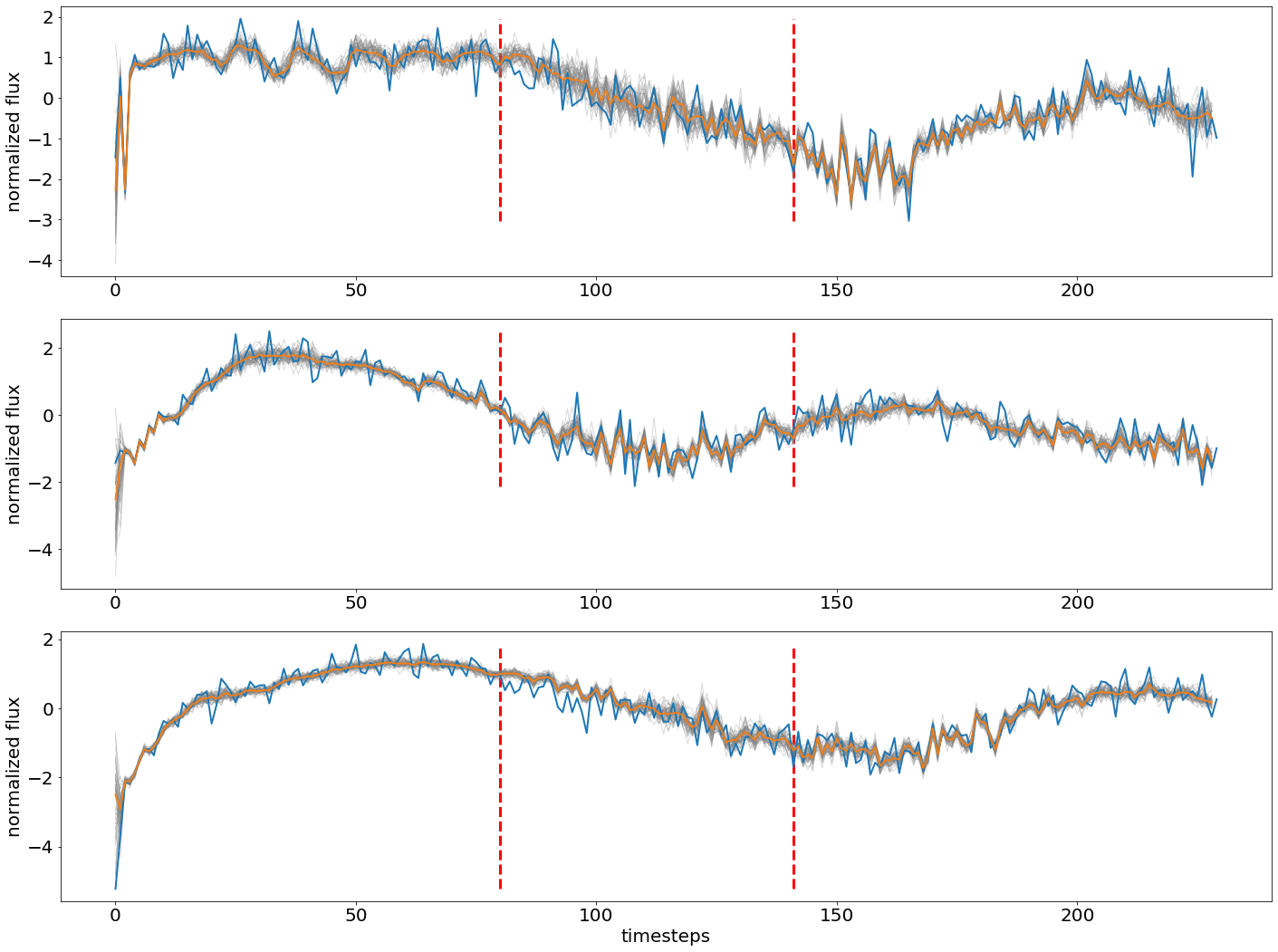}
\caption{Example of interpolations on 3 light curves containing no transit. The prediction range is located inside the vertical dashed lines. The raw light curve is displayed in blue, and the predicted traces in grey and the median prediction in orange.\label{fig:preds_test}}
\end{center}
\end{figure}

As the ground truth, i.e. the predicted stellar and instrumental flux, is not available in the transit range, we chose to first test the interpolating model on the pre-transit range instead, where its predictions can be evaluated more directly. In practice, three prediction ranges are selected in the first 250 timesteps of the time-series, where no transit signal is present, and the mean squared error (MSE) metric is used to evaluate and compare various models. An example of prediction is shown on Figure \ref{fig:preds_test}, where the prediction is obtained by averaging 50 sampled traces.

\subsubsection{Hyperparameter optimisation}
We perform a grid search over different types of inputs and hyperparameters. More specifically, we vary the aperture width of the sub-array used for computing the raw light curve between 5 and 7 pixels; we experiment with including and excluding covariate features, namely: 1) excluding covariate features altogether; 2) including centroid time-series, and 3) including centroid and PSF width time-series. Furthermore, we  vary the number of layers (between $1$ and $4$), units per layer (powers of $2$ up to $1024$ and dropout rate (between $0\%$ and $50\%$ in steps of $1\%$)~\footnote{Dropout is a common regularization technique in deep learning consisting in randomly reinitializing a fraction of the neurons of a given layer. The dropout rate refers to this fraction.} values for the LSTM block; and a unidirectional or bidirectional network\footnote{By `unidirectional' network we mean one that uses just past timesteps to infer the current one. With `bidirectional' we mean using timesteps from both past and future to infer the current one.}. We train each different model on the 6 light curves and 3 different prediction ranges, monitoring the average MSE for these 18 predictions and using it as a criterion for early stopping and comparison between the different models. 

From these tests we observe the following:
\begin{itemize}
    \item Including the centroids information improves the quality of the prediction by a factor of $\sim2$, and including the PSF widths time-series besides the centroids brings a further increase in MSE.
    \item Dropping $3\%$ of the recurrent units improves slightly the predictions, especially when the number of parameters of the network increases.
    \item Using a bidirectional network slightly decreases the quality of predictions.
\end{itemize}
More information on the hyperparameters and model training used is presented in Appendix \ref{sec:hyperparameters}.

\subsubsection{Performance}
We present in Table \ref{tab:perf_interp} the results of the best tested model in the explored grid. As a reference for the performance of the interpolation, we include a baseline model, which is a linear composition of the centroid X/Y time-series $\{z^X_t\}$ and $\{z^Y_t\}$:
$$ \widehat{y_t} = a + bz^X_t + cz^Y_t $$ where $a,b,c \in \mathbb{R}$. The model is trained on the training ranges\footnote{The model was fitted using scikit-learn's linear regression module: \url{https://scikit-learn.org/stable/modules/generated/sklearn.linear_model.LinearRegression.html}} and evaluated in the prediction ranges. The metrics computed for both models include the MSE, the mean-absolute error (MAE) and the mean signal-to-noise (SNR) ratio, defined as:
$$MSE = \frac{1}{N}\sum_{t_1}^{t_2}{(\widehat{y_t}  - y_t)^2.}$$
$$MAE = \frac{1}{N}\sum_{t_1}^{t_2}{|\widehat{y_t}  - y_t|.}$$
$$SNR = \frac{1}{N}\sum_{t_1}^{t_2}{|\widehat{y_t}  - y_t|/\sigma_{noise}.}$$
where N is the number of observations, and $\sigma_{noise}$ an estimate of the noise level computed by taking the mean value of the running standard deviation of width 15 over each input light curve.

\begin{table}
    \centering
    \begin{tabular}{c|c|c}
       LC instance & Baseline & This model \\
       \hline 
       MSE \#1 & 0.192 & 0.156\\   
       MSE \#2 & 0.782 & 0.196\\
       SEE \#3 & 0.606 & 0.138\\
       MSE \#4 &  0.0921 & 0.0529\\
       MSE \#5 & 0.275 & 0.249\\
       MSE \#6 & 0.688 & 0.0837\\
       Mean MSE &  0.439 & 0.124\\
       Mean MAE & 0.503 & 0.367 \\
       Mean SNR & 1.430 & 0.807 \\
    \end{tabular}
    \caption{Comparison of performance on the 6 light curves for each model. Every value is averaged between prediction and actual value over 3 different ranges of length $60$ and starting respectively at timeseteps $80$, $100$ and $120$. The three last lines show the mean performance over all light curves and ranges in terms of MSE, MAE and SNR.}
    \label{tab:perf_interp}
\end{table}

Given its simplicity, this baseline model does a remarkably good job at interpolating $\{x_t\}$, and this is why it was chosen here as a reference for the MSE. Furthermore, since the TLCD-LSTM also uses the centroid time-series, the increase in performance seen on Table \ref{tab:perf_interp} can directly be interpreted as the improvement brought by the LSTM's ability to identify temporal dependencies in the raw light curve.

\subsection{Prediction on real transit ranges}
\begin{figure*}
\begin{center}
\includegraphics[width=\textwidth]{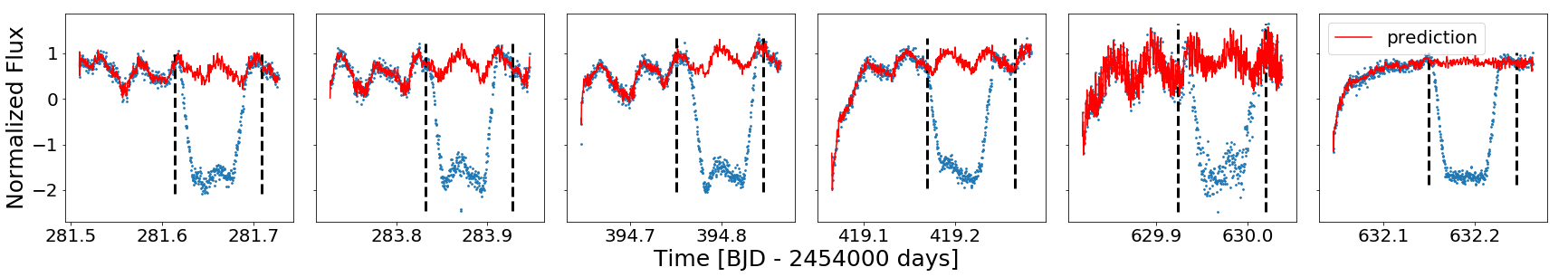}
\caption{Raw data (blue) and model output, i.e. interpolated light curve in the absence of transit (red) for the light curves. Dashed vertical lines indicate the initial prediction ranges. \label{fig:preds} }
\end{center}
\end{figure*}

Using the optimised hyperparameters listed in \ref{subsec:testing} and after training the model for 3000 epochs we extract the output of the network for the whole time ranges, shown in red on Figure \ref{fig:preds}. Note that the decreasing learning rate used guarantees the convergence of the network towards a stable solution. Visually, the model seems to be able to pick up the trends and variability of each time-series, while joining smoothly the pre and post transit ranges where the ground truth is known. 

\begin{figure*}
\begin{center}
\includegraphics[width=\textwidth]{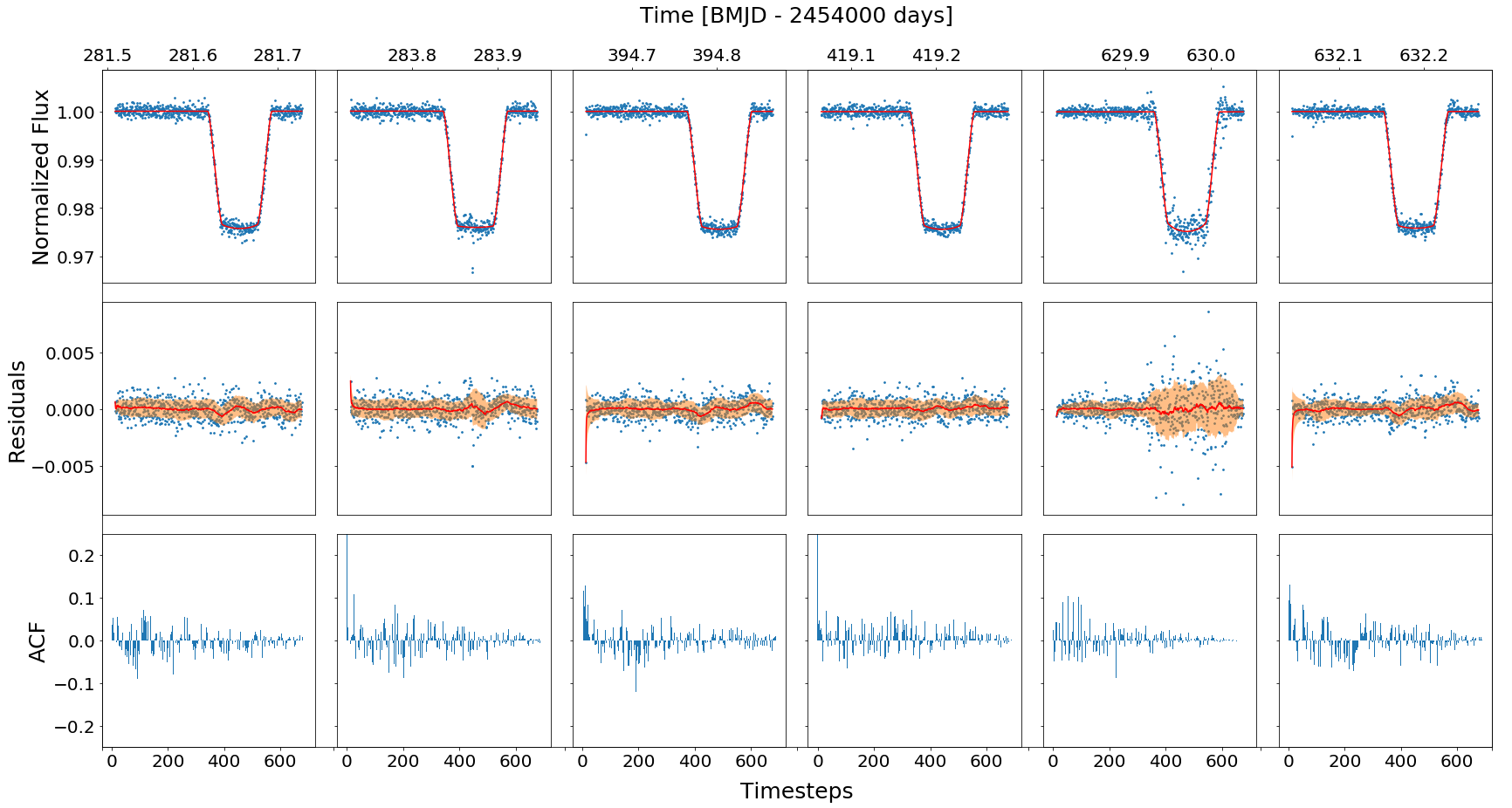}
\caption{\textit{(Top)} Best Transit Fit (red curve) to the detrended light curve (blue points) normalized with respect to the stellar flux. \textit{(Center)} Fit residuals (blue points) along with the moving average (red curve) and standard deviation ( orange) of the residuals. \textit{(Bottom)} Auto-Correlated Function of the residuals \label{fig:transit_fit} }
\end{center}
\end{figure*}

The last step is to perform the transit fit on the detrended light curve $\{1-\delta_t\}$ normalized with respect to the stellar flux $F_s(t)$. 
Since the limb darkening effect is minor at 8\,$\mu$m, we chose a transit model with linear limb darkening (bound between $0.05$ and $0.25$), and compute the best fit using a Markov Chain Monte Carlo optimization procedure\footnote{The transit model was fitted using PylightCurve package: \href{https://github.com/ucl-exoplanets/pylightcurve}{https://github.com/ucl-exoplanets/pylightcurve}} \citep{tsiaras_new_2016}. 
The fitted parameters are $R_p/R_s$, the mid-time transit time $t_c$, a linear limb darkening coefficient $u$, the orbit inclination $i$ and orbital semi-major axis relatively to the stellar radius $a/R_s$. The fitted model, residuals and auto-correlated functions (ACF) are shown in Figure\,\ref{fig:transit_fit} and the fitted parameters are presented in Table\,\ref{tab:transit_pars}. The higher variance present in the residuals of the $5th$ lightcurve is  due to a higher noise level in the input data for this light curve.

\begin{table*}
    \centering
    \begin{tabular}{c|c|c|c|c}
        \hline
       $t_c$ & $R_P/R_S$ & $i$ & $a/R_S$ & $u$ \\ 
        (BJD-2454000) &   & ($deg$) & & \\
       \hline 
		$281.655329\pm{0.000046}$ & $0.15489\pm{0.00018}$ & $85.7682\pm{0.0502}$ & $8.971\pm{0.045}$ & $0.141\pm{0.020}$ \\
		$283.873934\pm{0.000049}$ & $0.15477\pm{0.00024}$ & $85.7277\pm{0.0698}$ & $8.901\pm{0.069}$ & $0.051\pm{0.018}$ \\
		$394.802829\pm{0.000045}$ & $0.15564\pm{0.00020}$ & $85.5926\pm{0.0771}$ & $8.799\pm{0.065}$ & $0.093\pm{0.037}$ \\
		$419.206955\pm{0.000070}$ & $0.15520\pm{0.00015}$ & $85.8120\pm{0.0881}$ & $8.992\pm{0.075}$ & $0.129\pm{0.026}$ \\
		$629.971770\pm{0.000097}$ & $0.15523\pm{0.00042}$ & $85.9760\pm{0.1263}$ & $8.999\pm{0.106}$ & $0.248\pm{0.028}$ \\
		$632.190498\pm{0.000046}$ & $0.15488\pm{0.00019}$ & $85.5862\pm{0.0747}$ & $8.782\pm{0.057}$ & $0.097\pm{0.028}$ \\
		\hline
    \end{tabular}
    \caption{Fitted physical parameters for each of the 6 transits. \label{tab:transit_pars}}
\end{table*}

\begin{figure}
\begin{center}
\includegraphics[width=\columnwidth]{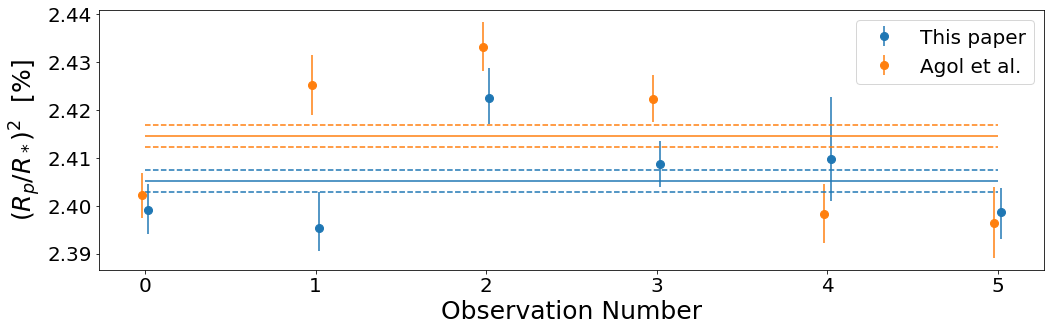}
\caption{Comparison of fitted transit depths between this work and \citet{agol_climate_2010}, for the six transit observations of HD189733b. The horizontal lines show the means of the observations from both papers weighted by their respective standard deviations. The dotted lines show the standard deviations of these weighted means. \label{fig:compar}}
\end{center}
\end{figure}

We compare the retrieved transit depths with the results published in \citet{agol_climate_2010} for the same data set and preprocessing steps (Figure \ref{fig:compar}). Although slightly smaller, the scatter of the predictions is still present with a standard deviation of $91.7$ppm instead of $144$ppm. The mean weighted by the standard deviations of the 6 transit depths is also found to be slightly smaller in our case by $94$ppm $\approx 4\sigma$.

\section{Discussion and conclusions} \label{sec:conclusion}

We presented a deep learning model suitable for interpolating time-series, and showed how it can be used to predict the variability of stellar light curves for subsequent transit fit. This approach has the advantage of not making any assumption on the types of noise, systematics or transit shape. 

The presented method is similar to the Gaussian Process (GP) approach (\citealp{gibson_gaussian_2012}, \citealp{rasmussen_gaussian_2005}) in that they both construct highly non-linear models, avoid explicit physical modelling of the systematics and provide probabilistic predictions. However, they differ in various aspects:

1) The neural network lightcurve interpolation approach we propose does not need any transit model whereas it is included in the kernel of the GP. This makes the TLCD-LSTM approach more generally applicable as it does not depend on a pre-defined kernel function. 


2) The GP approach requires fewer parameters to train and provide fully Bayesian predictions compared to our LSTM-based approach. The smaller number of free parameters may make GPs the preferred choice for short time series. However, GPs computation scales more poorly with the number of data points, preventing them to be applicable to datasets of more than $\approx1000$ time steps without binning of the time series. The proposed interpolating LSTM can on the other hand be applied to longer or multiple light curves as commonly found in Kepler and TESS time series allowing for even very long period variability to be captured in the predictive LSTM model. This is because the computational complexity in the case of GPs mainly depends on the number of data points, while in the case of the deep neural networks in the architecture chosen (i.e. the number of layers, number of nodes per layer \& type of layers in our case).

While the current implementation still relies on a few preprocessing steps such as computing the raw light curve or centroids fits, it constitutes a first step towards the ultimate goal of developing an end-to-end detrending pipeline where the input would be the raw pixel light curves or focal plane images. Furthermore, while we trained our network on data from six real light curves only, taking advantage of a large number of light curves, real or simulated, would allow developing a more general detrending approach for each instrument. LSTMs allow for efficient transfer learning between data sets and instruments (e.g. Kepler to TESS). This may become important in modelling common systematics such as stellar noise between planet-star systems observed by multiple instruments. 

As we have firmly entered the era of `big data' in planet detection (e.g. Kepler, TESS and ground based surveys) and with upcoming characterisation missions and instruments (e.g. JWST, Ariel, CHEOPS and the ELTs), the opportunities for data detrending and modelling with scalable deep learning methods, capable of processing large numbers of high dimensional data will become increasingly prevalent in the future.\\


\noindent\textbf{Software}

The data and code used in this paper are available on GitHub under a Creative Commons Attribution 4.0 International License (https://github.com/ucl-exoplanets/deepARTransit, archived on Zenodo) and a MIT License (https://github.com/ucl-exoplanets/pylightcurve).\\

\noindent\textbf{Acknowledgements}

This project has received funding from the European Research Council (ERC) under the European Union's Horizon 2020 research and innovation programme (grant agreement No 758892, ExoAI), under the European Union's Seventh Framework Programme (FP7/2007-2013)/ ERC grant agreement numbers 617119 (ExoLights) and the European Union's Horizon 2020 COMPET programme (grant agreement No 776403, ExoplANETS A). Furthermore, we acknowledge funding by the Science and Technology Funding Council (STFC) grants: ST/K502406/1, ST/P000282/1, ST/P002153/1 and ST/S002634/1.

\clearpage
\appendix
\section{Transit fit} \label{ap:transit_fit}
To obtain a light curve normalized with respect to the star, three steps are required: transformation to the original units $ y_t \rightarrow y'_t$, subtraction of background flux $F_b(t)$, and division of the background subtracted raw light curve by the predicted star flux: 

$$ 1-\delta(t) = \frac{F_{received}(t) - F_b(t)}{F_s(t)-F_b(t)} $$
With the time-series notations where $x'_t$ and $\hat{y}'_t$ are the input and mean prediction of the neural network in the original units: 
$$ 1-\delta_t = \frac{x'_t - F_{b,t}}{\hat{y}'_t - F_{b,t}}$$.

Note that during training, we use a simple piecewise-linear transit model with four parameters described in \citet{carter_analytic_2008} optimized by least-square fitting and neglect the contribution of the background $F_b \ll x'_t, \hat{y}'_t$. 

\section{Pre-processing} \label{ap:preprocessing}

Here we describe the different preprocessing steps applied to the raw subarray data.
\paragraph{Outlier removal}
Due to a number of causes, such as remaining cosmic rays or bad pixels, the flux on individual pixels can exhibit great fluctuations within short timescales ($\approx 1sec$). These abnormal values are identified by computing the absolute difference of the pixels' flux with their corresponding median within a time window of width 5 ($2\sec$ exposure). The values of the median-subtracted time-series greater than 4$\sigma$ are then replaced by the median values, where $\sigma$ is the standard deviation of the time-series. 

\paragraph{Raw light curve extraction}
In order to limit the influence of background light and focus on the brightest pixels of the stellar PSF, $3\times3$, $5\times5$ and $7\times7$ pixel regions are extracted around the brightest pixel.
The raw light curve is then obtained by summing all the individual pixel light curves. 

\paragraph{Centroid fitting}
As mentioned earlier the centroid position time-series are highly correlated with the flux received by the detector. In order to compute the centroids, we perform a two-dimensional Gaussian fit with offset to the data at every timestep, and hence extract four useful time-series, two of which are monitoring the position of the center on the detector and two for the width of the Gaussian. As discussed in \citet{agol_climate_2010}, this method provides by far a better estimate of the centroids over other methods such as the flux-weighted ratio extraction. 

\paragraph{Background extraction}
The background flux contribution to the total flux, although minor, increases with the aperture size used for the light curve extraction. We estimate it here by taking the median flux value of the pixels located in the four corners of each frame, corners delimited by the complement of a circular aperture of radius $16$. It accounts for $0.67\%$ to $1.2\%$ in our analysis, and should therefore be taken into account. However, as the background estimation is necessarily approximate, we advocate to still interpolate on the raw light curve directly, and only correct for it before the transit fit.

\paragraph{Normalization} The raw light curve and centroid time-series are all locally standardized, i.e. individually centered around a mean value of zero and rescaled to have their standard deviation equal to one.

The preprocessed raw light curves and centroid X/Y positions are shown on Figure \ref{raw_data}. Note the diversity of effects among them, showing more or less stochastic noise, ramps or jitter.

\section{Hyperparameters} \label{sec:hyperparameters}
\paragraph{Training parameters}
Training was performed using the ADAM optimizer \citep{kingma_adam:_2014} with parameter values $\beta_1 =0.9$, $\beta_2 =0.99$, $\epsilon=10−8$. The learning rate was decreased from 0.01 to 0.0001 using a polynomial decay law with exponent $20$. We train the model using a batch size of $6$ (all the lightcurves) for faster training.

\begin{table}[!hbt]
    \centering
    \begin{tabular}{c|c}
        Parameter & Value  \\
        \hline
        Number of LSTM layers & 3\\
        Number of units per layer & 256 \\
        Recurrent drop-out rate & 3 \% \\
        Initial bias values & 0.0 \\
        Batch size & 6 \\
    \end{tabular}
    \caption{Table of the network hyperparameters}
    \label{tab:hyperparams}
\end{table}

\section{Plots of fitted transit parameters}

\begin{figure*}[h]
\begin{center}
\includegraphics[width=\columnwidth]{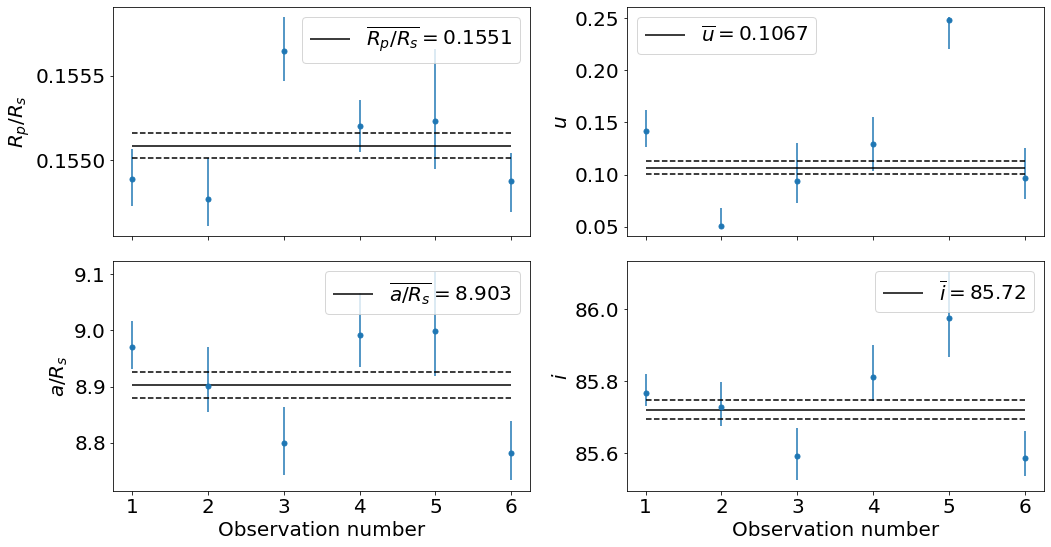}
\caption{Additional plots showing the fitted parameters $R_p/R_s$, $u$, $a/R_s$ and $i$ for each of the 6 light curves, as well as their weighted mean and associated standard deviation.}
\end{center}
\end{figure*}


\nocite{morvan_deepartransit_2019}
\bibliography{paper_0}{}
\bibliographystyle{aasjournal}



\end{document}